\documentclass{article}

\usepackage{graphicx}
\usepackage{epsfig}
\usepackage{amsfonts}

\usepackage{color} 
\usepackage[normalem]{ulem}
\usepackage{comment}

 \usepackage{amssymb}
\usepackage{amsmath}
\usepackage{hyperref}
\usepackage{graphicx}       
\usepackage{dcolumn}        
\usepackage{bm}             
\usepackage{amssymb}
\usepackage{amstext}

\date{\today}

\newcommand{\insertplot}[5]{\begin{figure}
 \hfill\hbox to 0.05in{\vbox to #5in{\vfill
 \inputplot{#1}{#4}{#5}}\hfill}
 \hfill\vspace{-.1in}
 \caption{#2}\label{#3}
 \end{figure}}
 \newcommand{\inputplot}[3]{
 \special{ps: plotfile #1}
}
\newcounter{fig}

\newcommand{\ee}{\end{equation}}
\newcommand{\eea}{\end{eqnarray}}

\newcommand{\beq}{\begin{equation}}
\newcommand{\eeq}{\end{equation}}

\begin{document}

\title{
\Large{\bf 
Are slowly rotating Ellis-Bronnikov wormholes stable?
}
}
 \vspace{1.5truecm}

\author{
{\large }
{ Bahareh Azad}$^{1}$\footnote{bahareh.azad@uni-oldenburg.de}, 
{ Jose Luis Bl\'azquez-Salcedo}$^{2}$\footnote{jlblaz01@ucm.es},
{ Fech Scen Khoo}$^{1}$\footnote{fech.scen.khoo@uni-oldenburg.de},
and
{ Jutta Kunz}$^{1}$\footnote{jutta.kunz@uni-oldenburg.de}
\\
\\
$^{1}${\small  Institut f\"ur  Physik, Universit\"at Oldenburg, Postfach 2503,
D-26111 Oldenburg, Germany}
\\
$^{2}${\small  Departamento de F\'isica Te\'orica and IPARCOS, 
Facultad de Ciencias F\'isicas, Universidad Complutense de Madrid, Spain} 
}

\maketitle

\begin{abstract}
We investigate the radial perturbations of Ellis-Bronnikov wormholes ($\mathrm{l}=0$) in a slowly rotating background expanded up to second order in rotation.
We find indications that simple wormhole solutions such as Ellis-Bronnikov in General Relativity can be stabilized by rotation, thus favoring a viable traversable wormhole. 
This opens up the intriguing question whether the many other wormhole solutions with or without the support of exotic matter can become linearly mode stable when the wormhole rotates.
\end{abstract}


\section{Introduction}

Wormholes have received much attention in recent years, in particular, concerning their possible astrophysical signatures such as gravitational lensing effects of wormholes
\cite{Cramer:1994qj,Safonova:2001vz,Perlick:2003vg,Nandi:2006ds,Abe:2010ap,Toki:2011zu,Nakajima:2012pu,Tsukamoto:2012xs,Kuhfittig:2013hva,Bambi:2013nla,Takahashi:2013jqa,Tsukamoto:2016zdu},
wormhole shadows
\cite{Bambi:2013nla,Nedkova:2013msa,Ohgami:2015nra,Shaikh:2018kfv,Gyulchev:2018fmd,Guerrero:2022qkh},
or wormhole accretion disks and quasi-periodic oscillations
\cite{Harko:2008vy,Harko:2009xf,Bambi:2013jda,Zhou:2016koy,Lamy:2018zvj,Deligianni:2021ecz,Deligianni:2021hwt}.
In fact, wormholes may mimic many properties of black holes 
\cite{Damour:2007ap,Bambi:2013nla,Azreg-Ainou:2014dwa,Dzhunushaliev:2016ylj,Cardoso:2016rao,Konoplya:2016hmd,Nandi:2016uzg,Bueno:2017hyj,Blazquez-Salcedo:2018ipc,Azad:2020ajs,Azad:2022qqn}.
Therefore the study of gravitational waves emitted by wormholes is also of great interest.

The observations of gravitational waves by the LIGO/VIRGO collaboration have provided us with invaluable information about the merger of highly compact objects such as black holes and neutron stars (see e.g. \cite{LIGOScientific:2016aoc,LIGOScientific:2017vwq}). 
Here the ringdown phase after the merger events is dominated by the characteristic frequencies of the final objects, encoded in their quasi-normal modes (QNMs).
Whereas the QNMs of black holes and neutron stars are well studied in General Relativity (GR) \cite{Kokkotas:1999bd,Berti:2009kk,Konoplya:2011qq}, the corresponding modes of wormholes have received much less attention. %

In GR classical wormholes need exotic matter in order to violate the energy conditions \cite{Ellis:1973yv,Ellis:1979bh,Bronnikov:1973fh,Morris:1988tu,Morris:1988cz,Visser:1995cc,Hochberg:1996ee,Hochberg:1998ii,Visser:2003yf,Bronnikov:2005gm,Alcubierre:2017pqm}.
While this issue can be circumvented in several ways (for example, by considering wormholes in generalized gravities \cite{Kanti:2011jz,Alcubierre:2017pqm} or by adding fermionic matter \cite{Blazquez-Salcedo:2020czn,Konoplya:2021hsm}),
another question concerns the stability of wormholes in GR and generalized gravities. 
For instance, the notorious radial instability of the static spherically symmetric Ellis-Bronnikov (EB) wormholes of GR has been known for long \cite{Shinkai:2002gv,Gonzalez:2008wd,Gonzalez:2008xk,Cremona:2018wkj,Blazquez-Salcedo:2018ipc}.

Here we present intriguing evidence that this radial instability of EB wormholes will disappear, when these wormholes rotate sufficiently fast.
However, unlike their static counterparts, there are no known analytical solutions of rotating EB wormholes \cite{Volkov:2021blw}, these are known only perturbatively for slow rotation \cite{Kashargin:2007mm,Kashargin:2008pk} and numerically for fast rotation \cite{Kleihaus:2014dla,Chew:2016epf}.
Additionally, rotating EB wormholes have been investigated in five dimensions for the case of two equal magnitude angular momenta, where the angular coordinates factorize \cite{Dzhunushaliev:2013jja}.

In four dimensions, a mode analysis of rotating compact objects becomes a challenging task.
Already for Kerr-Newman black holes linear mode stability could only be provided by solving the resulting system of coupled partial differential equations with sophisticated numerical methods \cite{Dias:2015wqa}.
Alternatively, the modes have recently been obtained perturbatively to second order in rotation, where parity mixing complicates their evaluation as opposed to first order \cite{Blazquez-Salcedo:2022eik}.

We here apply this recently developed second order formalism \cite{Blazquez-Salcedo:2022eik} in order to study the change of the unstable radial mode of EB wormholes in the presence of slow rotation.
We demonstrate, that the magnitude of the pertinent unstable eigenvalue decreases with increasing angular momentum of the wormhole, thus diminishing the instability.
In fact, in our perturbative analysis the eigenvalue 
{vanishes}
at some critical value of the angular momentum, that slightly depends on the asymmetry parameter of the wormhole, {and we argue that nonperturbatively the instability will disappear slightly 
earlier}.

\section{EB Wormholes in Slow Rotation}

We consider a phantom scalar field $\Phi$ minimally coupled to GR,
\begin{eqnarray}
			S &=& \frac{1}{16 \pi G}\int d^4x \sqrt{-g} 
		\Big[\mathrm{R} + 2 \partial_\mu \Phi \, \partial^\mu \Phi 
		 \Big] \, 
   \label{eq:ellis} 
\end{eqnarray}
with field equations %
\begin{eqnarray}
    \mathrm{R}_{\mu\nu} = -2 \partial_{\mu}\Phi\partial_{\nu}\Phi \, , \quad 
    \partial_{\mu}\partial^{\mu}\Phi = 0 \, .
\end{eqnarray}
Up to second order in rotation the background metric is given by
\begin{eqnarray}
ds^2 &=& -e^{f}\left[1+\epsilon_r^2 2\left(h_0(r)+h_2(r)P_2(\theta)\right)\right]dt^2 
+ e^{-f}\left[1+\epsilon_r^2 2\left(b_0(r)+b_2(r)P_2(\theta)\right)\right]dr^2 \nonumber \\ 
&+& 
{e^{-f}} R^2
\left[1+\epsilon_r^2 2\left(k_0(r)+k_2(r)P_2(\theta)\right)\right]
\times 
\left[
d\theta^2+\sin^2{(\theta)}\left[d\varphi-\epsilon_rw(r)dt\right]^2 
\right]
\ ,
\label{metric_1}
\end{eqnarray}
where $R^2=(r^2+r_0^2)$, with slow-rotation parameter $\epsilon_r \ll 1$, and the Legendre polynomial $P_2(\theta) = \left(3\cos^2{(\theta)}-1\right)/2$.
The corresponding second order phantom scalar field has the form
\begin{eqnarray}
\Phi = \phi(r) + \epsilon_r^2 \left(\phi_{20}(r)+\phi_{22}(r)P_2(\theta)\right) \ .
\label{scalar_1}
\end{eqnarray}
In the following we choose the gauge $k_0(r)=0$.
The throat of the wormhole is located at $r=0$, and it has an area {$\mathrm{A}=4\pi r_0^2 e^{\frac{\pi C}{2r_0}}$.}

The EB wormhole background solutions are
\begin{eqnarray}
    f(r) = \frac{{C}}{r_0} \left(\tan^{-1} \! \left(\frac{r}{r_0}\right)-\frac{\pi}{2}\right) \ , \ \phi(r)= \frac{Q_0 f}{C} \ ,
\end{eqnarray}
with asymmetry parameter $C=2M_0$, related to the static background wormhole mass $M_0$, and the static background phantom scalar charge %
{$Q_{0} = \sqrt{C^2/4 + r_0^2}$.}
Here we will focus on $C\ge 0$.
At the first order in rotation, we solve to obtain
\begin{eqnarray}
    w(r)= \frac{3J}{2C(C^2+r_0^2)}
    \left[
    1 - \left(
    1+2C\frac{C+r}{R^2}
    \right)e^{2f}
    \right] \, . \, \, \, \, \, \, \, \,
\end{eqnarray}
In the system of field equations, the differential equations of the radial functions $h_0,{b_0},\phi_{20}$ are decoupled from those of ${h_2}, k_2, {b_2}, \phi_{22}$.
When focusing on radial perturbations, we only need the functions $\phi_{20}$, {$b_0$}, $h_0$.
For the simpler 
{symmetric}
background wormholes ($C=0$)
they are given by
\begin{eqnarray}
    \phi_{20}(r) &=& 
    \frac{3J^2}{rr_0^3}
    \left( 
    \phi + \frac{\pi}{2} - \frac{3r}{2r_0} - \frac{4}{\pi r_0} - \frac{3r^2+4r_0^2}{2R^2\phi}
    \right)
    \, , \nonumber \\
    b_0(r) &=& \frac{\Delta M}{r}
    \left(
    1-\phi\frac{r_0}{r}
    \right)
    +
    \frac{6J^2}{rr_0^3}
    \left(
    \frac{r_0}{r}+\phi
    \right)
    -\frac{3J^2}{r^2r_0^2}\phi^2
        \, , \nonumber \\
    h_0(r) &=&  \frac{3J^2}{r_0^4}
    \left(
    \phi^2 + \frac{\pi^2-8}{2\pi}\phi - \frac{r_0^2}{R^2}
    \right) \, .
\end{eqnarray}

Solutions for $C>0$ are more involved, and hence provided 
{elsewhere \cite{Azad:2023}}.
The static mass $M_0$ is corrected by
$\Delta M =3J^2(\pi^2-8)/(2\pi r_0^3)$
in the limit of $C=0$ when $r\rightarrow +\infty$.
Therefore, for %
$C=0$, the total mass of the wormhole is no longer zero. %
In fact, it receives a contribution from the rotation.
This echoes the finding in \cite{Kashargin:2008pk},
that there is no massless rotating {EB} wormhole.
The static scalar charge $Q_0$ is also corrected by $\Delta Q=-\Delta M$.
The mass {and scalar charge corrections for $C>0$ are} given
{elsewhere \cite{Azad:2023}}.

\section{Stability Analysis:
Radial Perturbations}

We linearly perturb the metric field $g$ and the phantom field $\Phi$ up to first order in the perturbation parameter $\epsilon_q$, in our slowly rotating ($sr$) background considered up to second order in rotation $\epsilon_r^2$.
The perturbed metric is then given by
\begin{eqnarray}
g_{\mu\nu} &=& g^{(sr)}_{\mu\nu} + \epsilon_q \delta h_{\mu\nu}(t,r,\theta,{\varphi}) \, , 
\end{eqnarray}
the perturbation of the metric being a combination of polar ($P$) and axial ($A$) perturbations:
$ \delta h_{\mu\nu}=\delta h^{(A)}_{\mu\nu} + \delta h^{(P)}_{\mu\nu}$.
The perturbed phantom field is
\begin{eqnarray}
\Phi &=& {\Phi}^{(sr)} + \epsilon_q \delta{\phi}^{(P)} (t,r,\theta,{\varphi})
\, .
\end{eqnarray}
These perturbations can be decomposed in spherical harmonics 
$Y[\mathrm{l},\mathrm{m}](\theta,{\varphi})$ , 
with multipole numbers $\mathrm{l}$ and $\mathrm{m}$
\cite{Blazquez-Salcedo:2022eik}.
When wormholes rotate the perturbations of even (polar) and odd (axial) parities are entangled, i.e.~the multipole numbers $\mathrm{l}$ are coupled.
This results in an infinite tower of equations carrying the index $\mathrm{l}$ that are to be summed, as is well-known for other rotating objects such as black holes and neutron stars.
Here, we are interested in the radial-led perturbations, hence $\mathrm{l}=0$, $\mathrm{m}=0$ polar-led perturbations. 
For a given slowly rotating background up to second order in rotation, it is possible to show that these perturbations are just coupled to axial $\mathrm{l}=1$ perturbations ({see} \cite{Blazquez-Salcedo:2022eik}). Hence for our purposes, the metric perturbation can be simplified {to}
\begin{eqnarray}
\label{met_pert}
\delta h_{\mu\nu}= e^{i\omega t}
    \begin{pmatrix}
NY_0 & 0    & 0 & {S}_0 Y_\theta \\
0   & LY_0 & 0 & {S}_1 Y_\theta \\
0 & 0 & R^2 TY_0 & 0 \\
{S}_0 Y_\theta & {S}_1 Y_\theta & 0 & R^2 T\sin^2{\theta}Y_0
\end{pmatrix} \, \, \, \, \, \, \, \, \,
\end{eqnarray}
where 
$N, T, L, {S}_0, {S}_1$ are radial functions, 
$Y_0=Y[0,0]
=1/\sqrt{4\pi}$ 
and 
$Y_\theta=\sin{\theta} \, \partial_\theta Y[1,0]
=-\sin^2{\theta}\sqrt{3/4\pi}$. 
The phantom perturbation on the other hand can be taken as
\begin{eqnarray}
\label{phan_pert}
 \delta{\phi}^{(P)} = 
 {e^{i\omega t}}
 {\phi_1(r)} Y_0 \, .
\end{eqnarray}
With these perturbations, a consistent set of equations is formed, as the rotations are approximately slow and we {work} only up to second order in the rotation parameter $\epsilon_r$.


Following the procedure explained in  \cite{Blazquez-Salcedo:2022eik}, we obtain the linearized field equations.
For further simplification, we project the Einstein equation to the tensor spherical harmonics. In this way we get a set of polar and axial relations. 
Since we are interested only in radial-led $\mathrm{l}=0$ perturbations, it is enough for us to project the linearized field equations to the lowest tensor spherical harmonics, given by $Y[\mathrm{l}=0,\mathrm{m}=0]$.  This greatly simplifies the calculations as
most of the projected components of the field equations turn out to be zero.
Eventually
a system of ordinary differential equations (ODEs) for the radial functions $\phi_1, N, L$, $T$, $S_0$ and $S_1$ is formed.
However, these equations are not all independent. 

{After some manipulations,}
we are left with a system of four independent ODEs 
for the 6 perturbation functions $N, L, T, {S}_0, {S}_1$ and ${\phi_1}$ (found in equations  (\ref{met_pert})-(\ref{phan_pert})). Without loss of generality, and in order to solve the problem numerically, it is convenient to fix the gauge
by the choice $S_0=0$ and $\phi_1=0$. Then the linearized field equations reduce to a system of ODEs and algebraic relations, {for $C=0$}:

\begin{eqnarray}
\label{eqS1}
S_1(r) =
\frac{i\sqrt{3} J}{2 \omega r_0^3  R^2} 
\Big\{
R^{2} (r \pi -2 {r_0} ) L
+ 2  r_0^{3} N
-2 r R^{2}  \tan^{-1} (r/r_0)
(L +T )
+
\Big[\pi rR^2 -2r_0(r^2+2r_0^2)\Big] T
\Big\}
\, ,
\end{eqnarray} 

\begin{eqnarray}
\label{eqN}
\frac{dN}{dr} &=& 
-\frac{1}{{r R^2}}
\left[ (\omega^2R^4 +r_0^2 ) T + r_0^2 L \right]
+
\frac{J^2}{2\pi r^3r_0^6 R^4}
\Big\{ 
12 R^2 \tan^{-1}  
(r/r_0)
\Big[
-2\pi \omega^2 r r_0^3 R^4
\nonumber  \\ 
&-&
2\pi r r_0(r_0^4-3r^4-3r^2r_0^2)
+
 3\pi r^4R^2 
(\tan^{-1}  
(r/r_0)
- \pi)
+
r_0^4
\Big(\pi \tan^{-1}  
(r/r_0)
- (\pi^2+8)/2 \Big)
(\omega^2R^4+r_0^2)
  \nonumber  \\ 
&+& 3r
 2(\pi^2+8)r_0^7 \Big(\omega^2 (r_0^2 + 3r^2) +1\Big)
 + 6r^2r_0^5 
 ( - 2\pi^2 + 16/3)
 + 2r^4r_0^3 ( - 13\pi^2 + 8)
  \nonumber  \\ 
&+&
  2r^4r_0^3 \omega^2 (\pi^2+8) (r^2+3r_0^2)
  +  8 \pi r r_0^6 -12\pi^2r^6r_0
  + 3\pi^3 r^7
  + 6\pi r^5 r_0^2 (\pi^2+2)
  +3 \pi r^3 r_0^4 (\pi^2+8)
\Big]
T
  \nonumber  \\ 
&+&
  12 \pi  r^3 R^2 \tan^{-1}  
  (r/r_0)
  \Big[ 
  r_0(2r^2+r_0^2) 
 + r R^2 (\tan^{-1}  
  (r/r_0)
  - \pi) 
  \Big]
  L 
  + 6 r^3 r_0^3 R^2  (\pi^2-8) N
   \nonumber  \\     
&+& 3 r^3 R^2 \Big[r_0^3 (-3\pi^2+8)
    +\pi^3 r^3 
    + \pi rr_0\Big(r_0(\pi^2+4)-4\pi r\Big)
   \Big]
   L
\Big\}
\, ,
\end{eqnarray}

\begin{eqnarray}
\label{eqT}
	\frac{dT}{dr} &=&  
 -\frac{(L+T)r}{R^2}
  +
  \frac{J^2}{2\pi rr_0^3 R^4}
  \Big\{ 
12 \pi r R^2 \tan^{-1} (r/r_0) T 
    -
    3r \Big[(\pi^2+8)R^2-4\pi rr_0 \Big] T
    - 6rR^2 (\pi^2+8) L
 \nonumber \\  &+&
12 R^2 \tan^{-1} (r/r_0)
    \Big[ 2\pi (\pi r_0/4+r) + 4r_0
    - \, 
    \pi r_0 \tan^{-1} (r/r_0)
    \Big] L
  \Big\}
	\, ,
\end{eqnarray}

\begin{eqnarray}
\frac{dL}{dr} &=& 
\label{eqL}
\frac{1}{r R^2} \left[(\omega^2R^4+2r^2+r_0^2)T+(2r^2+r_0^2)L\right]
+
\frac{J^2}{2\pi r^3r_0^4 R^4}
\Big\{6R^2 \tan^{-1} (r/r_0)
    \Big[
     8\pi r r_0 R^2 (1+(\omega R)^2)
     \nonumber  \\  &+&
(\pi^2+8 -2\pi \tan^{-1} (r/r_0) )
  \Big[\omega^2 (2r_0^6+r^6)
 +   r_0^2 
      \Big(r_0^2(5\omega^2r^2+2) + r^2(4\omega^2r^2+3)\Big)\Big]  
    \Big]
    T
     \nonumber  \\ 
     &-& 12 r
     \Big[ \omega^2 r_0^7 (\pi^2+8)
     + \pi\omega^2 rr_0(r_0^5+4r^2r_0^3 
     +5r^4r_0)
       +  
      \pi rr_0^2 (3r_0^2+2r^2)
      + 2\pi\omega^2 r^7
       \nonumber  \\ 
       &+&  
       r_0 (\pi^2+8) \Big((\omega r)^2 \Big(3r_0^4+r^4+3r^2r_0^2\Big) + R^4\Big)
     \Big] T
      -
      12 r_0 R^2 \tan^{-1} (r/r_0)
       \Big[ (\pi^2+8) (r_0/2) R^2
      \nonumber  \\ 
      &+& 
      \pi r(2r_0^2+r^2)
      - \pi \tan^{-1} (r/r_0) r_0 R^2
      \Big] L 
       + 3 rr_0R^2
       \Big[ (\pi^2+8) (2r_0^2+r^2)
       + 4\pi rr_0
       \Big] L
\Big\}
\, ,
\end{eqnarray}

\noindent 
where $R^2=r^2+r_0^2$.
More technical details of our slow-rotation approximation framework can be found in \cite{Blazquez-Salcedo:2022eik}.


%
It can be shown that the 
{final}
system can be cast into a single second order ordinary differential equation for $T$, with algebraic relations for the rest of the perturbation functions. 
%
%

First note that equations (\ref{eqT}) and (\ref{eqL}) do not couple to the perturbation function $N$. By taking the derivative of equation (\ref{eqT}), and using again equations (\ref{eqT}) and (\ref{eqL}) to write $L$ and $dL/dr$ in terms of $T$ and $dT/dr$, the resulting second order equation has the form
\begin{eqnarray}
\label{eqd2T}
\frac{d^2T}{dr^2}+ A(r)\frac{dT}{dr} + B(r)T = 0 \, ,
\end{eqnarray}
where the coefficients $A,B$ are combinations of the coefficients of perturbation equations 
(\ref{eqT})-(\ref{eqL}). 
These coefficients depend on the background metric functions. 
For simplicity, it is convenient to write them like
\begin{eqnarray}
    A(r) = \frac{r_*'' }{r_*'}+2\frac{\alpha'}{\alpha} \, , \\
    B(r) = \frac{\alpha'' }{\alpha}-\frac{r_*''\alpha' }{r_*'\alpha}-2\left(\frac{\alpha'}{\alpha}\right)^2+(r_*' )^2 (V-\omega^2)  \, ,
\end{eqnarray}
that is, in terms of the tortoise coordinate $r_*$ and the amplitude $\alpha$, which are defined as follows:
\begin{eqnarray}
\label{tor_coord}
   \partial_r r_*
   = e^{-f}
   \left[
   1 +
   \epsilon_r^2 (e^{-f}b_0-h_0)
   \right] \, ,   \, \, \, \, \, \, \, \, \, \, \, \,  \\
   \frac{\partial_r\alpha}{\alpha}
   = \frac{(\partial_r\phi)^2R^2
   +\epsilon_r^2
   [
   2e^{-f}b_0
   -
   e^{-2f}R^4(\partial_r w)^2/6
   ]
   }{r-\sqrt{R^2(\partial_r\phi)^2-r_0^2}}
   \, .  \, \, \, \, \, \, \, \, \, \, \, \, 
\end{eqnarray}

The perturbation function $L$ can be expressed in terms of $T$ and $dT/dr$ from equation (\ref{eqT}), the function $N$ can be obtained from equation (\ref{eqN}), resulting in a relation $N=-L-2T+\mathcal{O}(J^2)$, and the remaining  function $S_1$ can be obtained from equation (\ref{eqS1}). Hence it is sufficient to solve the second order equation for $T$ in order to obtain the quasi-normal modes.
This equation can be written as a standard stationary Schrödinger equation with a potential $V(r)$. Finding the quasi-normal modes is 
thus reduced to an ordinary eigenvalue problem. 
%

If we take $T =\alpha Z$, the perturbation equation can be cast into a Schrödinger equation,
\begin{eqnarray}
\label{schroe}
    \partial^2_{r_*}Z
    + (\omega^2 - V(r))Z = 0 \, ,
\end{eqnarray}
where $\omega$ is the quasi-normal mode frequency, and $V(r)$ is the potential.
In this case, the potential is given by
\begin{eqnarray}
    V(r) &=& V_{static}(r) + J^2 \, 
    V_2(r) \, .
    \label{pot_term}
\end{eqnarray}
There is no linear term in the angular momentum $J$.
Its contribution only enters at second order,
and it is a function of the static background solutions $f$, $\phi$, 
the first order function $w$, and the second order functions $h_0$, $b_0$, $\phi_{20}$.
Taking $C$ to zero gives rise to the following potential
\begin{eqnarray}
\label{pot_c0}
&& V_{C=0}\,{(r)} =\frac{3r^2r_0^2+2r_0^4}{r^2R^4}+ 
\\
&&
J^2 
\Bigg(
\frac{-12\phi^2}{r^4R^2}+
 \frac{12\pi r
(R^4+r_0^4)
-6r_0^3
(\pi^2-8)
R^2}
{\pi r_0^3 r^4 R^4}\phi+
\frac{
3r(\pi^2-8)
(R^4+r_0^4)
+12\pi r_0
(R^2r^2+2r_0^4)
}{\pi r_0^3 r^4 R^4}
\Bigg) \, . \nonumber
\end{eqnarray}
We provide the complete expression of the potential term for $C>0$ %
{elsewhere \cite{Azad:2023}}.
In Figure \ref{Fig_pot}, we {show} %
the correction to the potential (\ref{pot_term}) of the perturbation system due to slow rotation, obtained with 
{symmetric}
($C=0$) and 
{asymmetric}
($C\ne 0$) background wormhole configurations.
The potential shown has been scaled with %
$\left(r-C/2\right)^2$.

\section{Numerical Method}

%
To obtain the unstable radial-led modes, one can start with an asymptotical analysis of equation (\ref{eqd2T}). Consider radial distances asymptotically far from the throat, meaning $|r|\to\infty$. By expanding the coefficients it is straightforward to find that at leading order $A(r)\to \mathcal{O}(1/r^2)$ and $B(r)\to -\omega^2$. This of course means that asymptotically far from the throat, our perturbations follow a standard wave equation in flat space-time. General solutions are given by a combination of outgoing and ingoing waves that behave asymptotically like $T(r) \approx e^{\pm i\omega r}$.

Here we are interested in unstable modes, {therefore} %
we focus on perturbations for which the eigenvalue $\omega$ is a purely imaginary number, $\omega = i\omega_I$, such that $\omega_I<0$. 
In this case, an asymptotic analysis of the perturbation function $T(r)$ reveals that it has to decay exponentially to zero at both infinities, $T(\pm\infty)=0$. 
To numerically solve the master equation and find the eigenvalue $\omega$, we compactify the radial coordinate so that $r = r_0 \tan(\pi x/2)$. 
In this compactified coordinate, the boundaries $x=\pm 1$ correspond to the two asymptotically flat infinities, $r=\pm \infty$, while at $x=0$ we have the wormhole throat, $r=0$.

In practice we solve the second order ODE for the function $T(r)$ together with 
an
auxiliary equation $E'=0$ where $E=\omega^2$. 
Non-trivial regular solutions are obtained by fixing the amplitude of the perturbation function to an arbitrary real constant at an arbitrary point $r=r_c$, i.e. $T(r_c)=constant$. 
The three boundary conditions we impose are $T(\pm1)=0$ and $T(x_0)=1$, where $x_0$ is taken as an arbitrary point in the domain of integration.

\begin{figure}[h!]
		\centering
    		\includegraphics[width=0.32\textwidth,angle=-90]{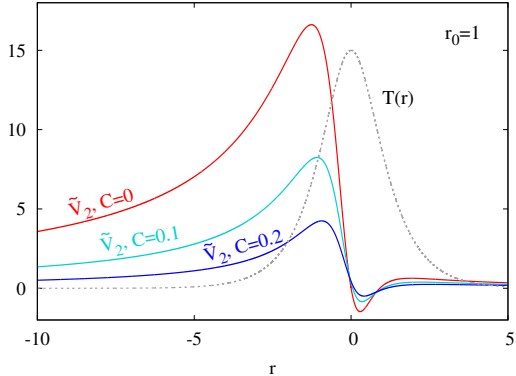}
		\caption{Potential correction $\tilde{V}_2=V_2(r-C/2)^2$ {vs radial coordinate $r$} for several values of $C$ for $r_0=1$, {and} %
  a typical profile of the perturbation function $T(r)$ {(in grey)}.} 
		\label{Fig_pot}
	\end{figure}

As an initial guess, we employ a standard Gaussian-like function for $T(x)$ that is zero at the boundaries, and a constant value for $E(x)$, which typically should be close to the true value of the eigenfrequency in order for the numerical ODE solver COLSYS \cite{Ascher:1979iha} to converge.
This ODE solver is designed for boundary value problems and implements a spline collocation method.
COLSYS automatically adapts the mesh points to achieve a prescribed precision of the functions and performs error estimates.
Typical estimated errors of the functions for the obtained unstable radial-led modes are on the order of $10^{-12}$.
This method has been used successfully before to generate unstable modes of the static EB wormholes \cite{Blazquez-Salcedo:2018ipc}
(and also applied to scalarized black holes in \cite{Blazquez-Salcedo:2018jnn}).
As an illustrative example, we show in Figure \ref{Fig_pot} a typical Gaussian profile of the function $T(r)$.
In order to generate the values of $\omega_I$, we start from the static limit ($J=0$), fixing $r_0=1$ and an arbitrary value of $C$. We reproduce the results from \cite{Blazquez-Salcedo:2018ipc}, and by slowly increasing the value of the angular momentum $J$, we get a relation
\begin{eqnarray}
    \omega_I = \omega_I^{(0)} + J^2 \Delta\omega_I^{(2)} \, ,
\end{eqnarray}
where $\omega_I^{(0)}$ is the unstable mode of the static EB wormhole, and $\Delta\omega_I^{(2)}$ is the slow rotation correction to the mode.
%

\section{Radial Instability}
\begin{figure}[h!]
		\centering
    		\includegraphics[width=0.32\textwidth,angle=-90]{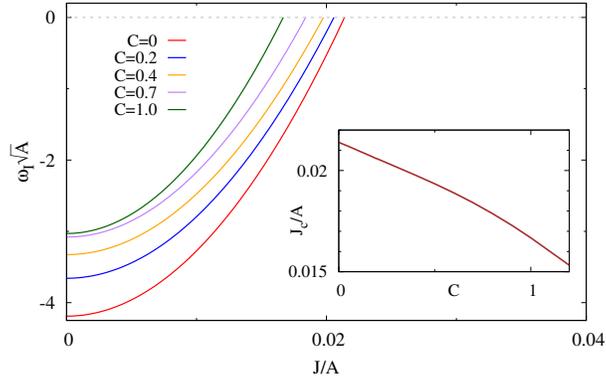}
		\caption{{Main} unstable mode {$\omega_I \sqrt{A}$ vs angular momentum $J/A$ (scaled with throat area $A$)} for several values of $C$. Inset: Critical {values $J_c/A$ vs $C$}.}  
		\label{Fig_J_omegaI_inset}
	\end{figure}

We exhibit in Figure \ref{Fig_J_omegaI_inset} the dependence of the unstable radial mode on the scaled and dimensionless angular momentum $J/A$ of the EB wormholes, where $A$ is the area of the throat (obtained in second order).
We recall that rotating EB wormholes approach the extremal Kerr black hole as their limiting configuration for large angular momenta \cite{Kleihaus:2014dla,Chew:2016epf}.
Thus the limiting case {here} is given {by} $J/A\to 1/(8\pi) \approx 0.0397$. Since {fast} rotating EB wormholes cannot exceed this limit, 
the slow rotation approximation {can} only make sense %
{below} this limit.

In Figure \ref{Fig_J_omegaI_inset} we show that, as we increase the value of the angular momentum, the value of $|\omega_I|$ decreases to zero. This is because, for every value of $C$ we have studied, the second order correction to the mode is positive, $\Delta\omega_I^{(2)}>0$. 
As a result, it is always possible to obtain a critical value of the angular momentum, $J_c$, for which the %
{eigenvalue reaches zero} %
and hence {this unstable mode}
disappears. 
Most importantly, {this happens}
at a {value of $J$} much smaller %
{than the} limiting value.
{In fact,} as seen in the inset of Figure \ref{Fig_J_omegaI_inset}, the 
critical values of the scaled angular momentum {$J_c/A$ decrease monotonically with C.}
{Moreover,} for each of the values of $C$ examined, 
{$J_c/A$} is around $50\%$ of the limiting value, which is well within the regime for which the second order slow rotation approximation has been shown to work {well} (for example, the quadratic approximation reproduces with excellent precision the spectrum of QNMs of Kerr-Newman black holes \cite{Blazquez-Salcedo:2022eik}).
{
In Table \ref{ImOm_coeff_main} are the values from the main branch of the unstable radial-led modes in Figure \ref{Fig_J_omegaI_inset}. 
Here $\omega^{(0)}_I$ corresponds to the unstable mode of the static Ellis wormhole, while $\Delta\omega^{(2)}_I$ is the second order correction in spin to the frequency. 
Note that $\Delta\omega^{(2)}_I$ is \emph{always} positive, showing the tendency of the angular momentum to stabilize the wormhole. 
We checked numerically that the 
first order correction
is compatible with zero, finding that it is always smaller than $10^{-7}$
(note that in general the linear correction is proportional to the angular number $\mathrm{m}$ (see e.g. \cite{Blazquez-Salcedo:2022eik}). Here $\mathrm{m}$ is zero).

\begin{table}[h!]
\begin{center}
\begin{tabular}{ || c | c c || }
 \hline
$C$ & $\omega^{(0)}_I$ & $\Delta\omega^{(2)}_I$ 
\\ 
  \hline
0 & -1.182 & 16.358
  \\ 
   0.05 & -1.094 & 13.356
  \\ 
   0.1 & -1.016 & 10.663
  \\ 
  0.2 & -0.882 & 7.036
  \\ 
  0.4 & -0.686 & 3.162
  \\ 
  0.7 & -0.501 & 1.040
  \\ 
  1.0 & -0.390 & 0.384
  \\ 
  \hline
\end{tabular}
\end{center}
\caption{Main unstable radial-led mode for several values of asymmetry parameter $C$.}
\label{ImOm_coeff_main}
\end{table}
}

  \begin{figure}[h!]
		\centering
    		\includegraphics[width=0.32\textwidth,angle=-90]{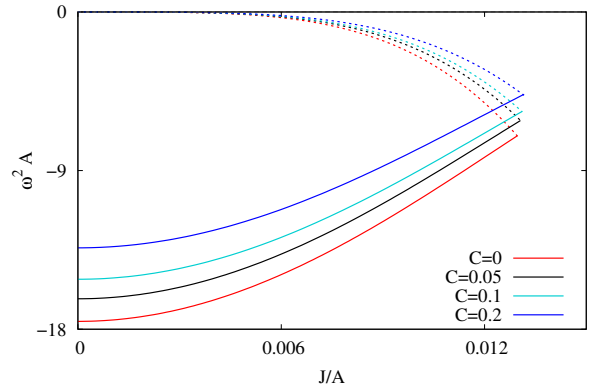}
		\caption{Main unstable mode {(solid)} %
  and second unstable mode {(dotted)
  vs angular momentum $J/A$ (scaled with throat area $A$)} 
  for $C=0$, $0.05$, $0.1$ and $0.2$.} 
		\label{Fig_J_w2_both}
	\end{figure}	

However, the critical {value} {of the} angular momentum
for which the wormholes {become stable} %
is presumably still %
smaller, as the following %
observations suggest.
When the wormholes are set into rotation, a zero mode of the static wormholes changes into a second unstable radial mode of the slowly rotating EB wormholes.
Following a quadratic dependence with a small negative coefficient $\Delta\omega_I^{(2)}$, this mode remains at first very small in magnitude.
Subsequently, its magnitude increases and it crosses the corresponding main unstable mode {not too far from} %
its critical {value $J_c/A$}. %
We exhibit both unstable modes in Figure \ref{Fig_J_w2_both}. 
Note that for the values of $C$ displayed the crossing occurs around $J/A\approx0.013$, slightly above $30\%$ of the limiting {value}. %
{
We show in Table \ref{ImOm_coeff_sec} the fitting coefficients for the second branch. Note that in the static limit the mode in this branch tends to a zero mode, and again there is no linear dependence.

\begin{table}[h!]
\begin{center}
\begin{tabular}{ || c | c  || }
 \hline
$C$ &  $\Delta\omega^{(2)}_I$ 
\\ 
  \hline
0  & -28.185
  \\ 
   0.05  & -21.431
  \\ 
   0.1 & -16.697
  \\ 
  0.2 & -10.220
  \\ 
  \hline
\end{tabular}
\end{center}
\caption{Second unstable radial-led mode for several values of asymmetry parameter $C$.}
\label{ImOm_coeff_sec}
\end{table}
}

We do not continue both modes beyond their crossing, however, since {this crossing} is indicating a %
{deficiency} of the applied quadratic approximation.
A non-perturbative study will allow for a far stronger dependence on the angular momentum than the current quadratic one.
We therefore conjecture, that when nearing the critical value, both unstable modes will change much more rapidly, approach each other, merge and disappear.
They will thus reflect the behavior observed in higher dimensions in the special case of wormholes with equal angular momenta \cite{Dzhunushaliev:2013jja}.

\section{Conclusions}

While EB wormholes represent rather simple wormhole solutions they may be considered prototypes whose properties can teach a lot about wormholes in general.
Therefore their response to rotation is highly remarkable.
Indeed, the potential of rotation to stabilize previously unstable wormholes was suggested before (see e.g. \cite{Matos:2005uh}), but we have given first evidence here, that it can indeed happen for wormholes in four space-time dimensions, where rotation implies significant loss of symmetry (associated with the technical challenge).

As we have shown above, the notorious unstable mode of static EB wormholes becomes more stable, as slow rotation is turned on.
In our second order and thus quadratic approximation the mode then becomes stable at a critical value of the angular momentum, which depends on the asymmetry parameter $C$.
For the EB wormholes we have considered, this critical value is around $50\%$ of the limiting angular momentum.
Moreover, we have observed that for rotating wormholes a second mode appears, {and} similar to the higher dimensional case \cite{Dzhunushaliev:2013jja}, it tends to merge with the fundamental unstable mode. 
Hence we have conjectured the behavior of the radial instability for the case of a future non-perturbative study: the radial instability will disappear at even smaller values of the angular momentum, around $30\%$ of the limiting {value}. 
Thus for {sufficiently} rapid rotation the 
wormholes will be stable.

\section*{Acknowledgements}

We gratefully acknowledge support by DAAD, the DFG Research Training Group 1620 \textit{Models of Gravity}, 
DFG project Ku612/18-1, 
FCT project PTDC/FIS-AST/3041/2020, COST Actions CA15117 and CA16104, and MICINN project PID2021-125617NB-I00 ``QuasiMode".
JLBS gratefully acknowledges support from Santander-UCM project PR44/21‐29910.

\begin{small}

 \end{small}

\end{document}